\documentclass[preprint2,times,tighten]{aastex61}
\usepackage{amsmath,amstext}
\usepackage[figure,figure*]{hypcap}
\usepackage{newtxmath} %use times font for math

\usepackage{txfonts}
\shorttitle{Breakthrough Listen: GBT Instrumentation}
\shortauthors{BL team}

\begin{document}

\title{The Breakthrough Listen Search for Intelligent Life: A Wideband Data Recorder System for the \\
       Robert C. Byrd Green Bank Telescope}

%\altaffiltext{1}{Department of Astronomy, University of California, Berkeley, 501 Campbell Hall \#3411, Berkeley, CA, 94720, USA}

\author{David H.E. MacMahon}
\affiliation{Radio Astronomy Laboratory, University of California at Berkeley}

\author{Danny C. Price}
\affiliation{Astronomy Department, University of California at Berkeley}

\author{Matthew Lebofsky}
\affiliation{Astronomy Department, University of California at Berkeley}

\author{Andrew P. V. Siemion}
\affiliation{Astronomy Department, University of California at Berkeley}
\affiliation{Radboud University, Nijmegen}
\affiliation{SETI Institute, Mountain View, California}

\author{Steve Croft}
\affiliation{Astronomy Department, University of California at Berkeley}

\author{David DeBoer}
\affiliation{Radio Astronomy Laboratory, University of California at Berkeley}

\author{J. Emilio Enriquez}
\affiliation{Astronomy Department, University of California at Berkeley}
\affiliation{Radboud University, Nijmegen}

\author{Vishal Gajjar}
\affiliation{Space Sciences Laboratory, University of California at Berkeley}

\author{Gregory Hellbourg}
\affiliation{Astronomy Department, University of California at Berkeley}

%\author{Jack Hickish}
%\affiliation{Radio Astronomy Laboratory, University of California at Berkeley}

\author{Howard Isaacson}
\affiliation{Astronomy Department, University of California at Berkeley}

\author{Dan Werthimer}
\affiliation{Space Sciences Laboratory, University of California at Berkeley}
\affiliation{Astronomy Department, University of California at Berkeley}
\affiliation{Radio Astronomy Laboratory, University of California at Berkeley}

\author{Zuhra Abdurashidova}
\affiliation{Radio Astronomy Laboratory, University of California at Berkeley}

\author{Marty Bloss}
\affiliation{Green Bank Observatory}

\author{Joe Brandt}
\affiliation{Green Bank Observatory}

\author{Ramon Creager}
\affiliation{Green Bank Observatory}

\author{John Ford}
\affiliation{Steward Observatory, University of Arizona}

\author{Ryan S. Lynch}
\affiliation{Green Bank Observatory}

\author{Ronald J. Maddalena}
\affiliation{Green Bank Observatory}

\author{Randy McCullough}
\affiliation{Green Bank Observatory}

\author{Jason Ray}
\affiliation{Green Bank Observatory}

\author{Mark Whitehead}
\affiliation{Green Bank Observatory}

\author{Dave Woody}
\affiliation{Green Bank Observatory}

\newcommand{\bw}{6~GHz }
\newcommand{\drate}{24~GB/s }
\newcommand{\dr}{DR}

% Helpers for referencing figures and sections etc
\newcommand{\refsec}[1]{Sec.~\ref{#1}}
\newcommand{\reffig}[1]{Fig.~\ref{#1}}
\newcommand{\reftab}[1]{Tab.~\ref{#1}}

\begin{abstract}
The Breakthrough Listen Initiative is undertaking a comprehensive search for radio and optical signatures from extraterrestrial civilizations.  An integral component of the project is the design and implementation of wide-bandwidth data recorder and signal processing systems.  The capabilities of these systems, particularly at radio frequencies, directly determine survey speed; further, given a fixed observing time and spectral coverage, they determine sensitivity as well. Here, we detail the Breakthrough Listen wide-bandwidth data recording system deployed at the 100-m aperture Robert C. Byrd Green Bank Telescope. The system digitizes up to \bw of bandwidth at 8 bits for both polarizations, storing the resultant \drate of data to disk.  This system is among the highest data rate baseband recording systems in use in radio astronomy.  A future system expansion will double recording capacity, to achieve a total Nyquist bandwidth of 12 GHz in two polarizations. In this paper, we present details of the system architecture, along with salient configuration and disk-write optimizations used to achieve high-throughput data capture on commodity compute servers and consumer-class hard disk drives.  \end{abstract}


\section{Introduction}
The search for extraterrestrial intelligence\footnote{often referred to by the acronym {\em SETI}} represents one of the  three primary means by which we may eventually determine whether life exists, or has ever existed, independent of the Earth.  Compared with the other two techniques: in-situ sampling, and remote sensing of exoplanet atmospheres and surfaces for signs of simple biology,  searches  for intelligent life have two unique advantages.  First, SETI is unique in probing for life that is at least as advanced as human beings -- and thus has the potential to answer a deeper question about the prevalence of conscious, technologically capable life in the universe.  Second, they are capable of probing a much larger volume of space than other search programs. Emission from our own technologies would be detectable out to {\em at least} several kpc by a civilization with 21st-century-Earth-scale telescopes, given sufficient light travel time.  This is  millions of times the volume  that might be probed in ground or space-based searches for disequilibrium chemistry in exoplanet atmospheres, surface indicators of  basic life on exoplanet surfaces, or in-situ  investigations of bodies in our own or nearby planetary systems.

The beginning of the modern search for extraterrestrial intelligence is generally marked by the publication of two seminal papers, \cite{Cocconi1959} and \cite{Drake1961}.  The first laid out key advantages of narrow-band radio signals; pointing out the relative ease at which cm-wave radio signals propagate through the interstellar medium and the earths atmosphere, and the energy efficiency of narrow band transmissions.  The latter described the first systematic search for narrow-band radio signals using a custom built analog receiving system attached to an 85-ft radio telescope at the newly-opened National Radio Astronomy Observatory (NRAO) in Green Bank, WV.

Decades of subsequent work clarified and extended these early ideas and experiments to include a wider range of radio frequencies and signal types, different parts of the electromagnetic spectrum,  indirect emission from technology unrelated to communication, and  wholly-different information carriers (e.g. neutrinos).  An excellent review of the state of the field up to 2003 is provided by \cite{Tarter:2003p266}.

\subsection{Digital Instrumentation}
Despite replete suggestions in the literature regarding so-called ``magic'' frequencies at which to conduct searches for narrow-band radio emission, e.g. \cite{1993ASPC...47..161G, 1996A&A...306..141M},  we have little a priori reason to believe one portion of the electromagnetic spectrum might be preferred. 
We are left with selecting frequencies based  on experimental practicality: it is cheaper to do radio astronomy on the ground than in space; from the ground, atmospheric opacity places limits on observable frequencies; extinction (i.e. signal attenuation) during propagation through the interstellar medium is not significant at radio wavelengths. Ultimately, these constraints leave us with several hundred GHz of radio spectrum to explore.  This fact drives the development of  ever-wider bandwidth search systems, enabling greater sensitivity for a fixed observing time and desired total searched bandwidth.

The burgeoning semiconductor and microelectronics industries were identified early on as a boon to radio astronomy, and particularly to searches for intelligent life \citep{NASA:2003p185,Morrison:1977p182}.  When analog signals from radio telescopes are digitized and processed electronically, the number of radio channels searched depends only on the speed of the electronics.  As computer technology has grown exponentially, so too has the power of instrumentation used in radio searches for extraterrestrial intelligence.  Several decades of the SERENDIP\footnote{Search for Extraterrestrial Radio Emissions from Nearby Developed Intelligent Populations} program \citep{Werthimer1985, Werthimer1995, Cobb2000, Kondo:2010p3277, Chennamangalam2017}, the development of the Allen Telescope Array \citep{Welch2009} exemplify this trend.  Searches for technological emission with the Murchison Widefield Array \cite{Tingay:2016go}, and planned searches with the Square Kilometre Array \cite{2015aska.confE.116S} and future mid-frequency aperture arrays \cite{2016arXiv161207917V} further illustrate the significant gains in experimental capabilities afforded by modern electronics.

\subsection{Breakthrough Listen}
The Breakthrough Listen Initiative (BL) was announced in July 2015 with a goal of conducting the most comprehensive, sensitive and intensive search for extraterrestrial intelligence in history.  Breakthrough Listen has contracted observing time with three  major research-class telescopes; two radio facilities, the Green Bank Telescope (GBT) in the northern hemisphere and the Parkes Telescope in the southern hemisphere, and one optical facility, the Automated Planet Finder (APF) at Lick Observatory in Northern California.

With these three facilities, BL has access to unprecedented amounts of primary observing time on world-class radio telescopes.  This is particularly distinct from the bulk of previous radio searches for extraterrestrial intelligence, which have been conducted commensally, by ``piggy-backing" on other observational programs.  With dedicated time, the Breakthrough Listen Team can perform targeted observations of target sources of their own choosing, control spectral coverage, and control observing cadences \citep{Isaacson2017}.  These capabilities are crucial for placing distinct and describable limits on the prevalence of extraterrestrial technologies in the event of a non-detection \citep{EnriquezBL2017}.  In order to maximize the utility of valuable observing time on the GBT and Parkes telescope, BL is deploying wide bandwidth digital instrumentation to these telescopes to record and analyze as much as of the available instantaneous bandwidth as possible. \\

The remainder of this paper provides an overview of the BL digital recorder system (henceforth \dr) for the Green Bank Telescope, and is structured as follows:  In \refsec{sec:architecture}, an architectural overview of the system from telescope to data disks is presented.  \refsec{sec:deployment} describes the deployment strategy.  More detailed discussions of the various hardware and software components of the system are described in \refsec{sec:hardware}, \refsec{sec:firmware} and \refsec{sec:software}, respectively.  \refsec{sec:integration} describes the integration of the BL \dr ~with the GBT's existing monitor and control system: a critically important aspect of deploying ``third-party" back-ends on observatories such as the GBT.  System verification and initial on-sky results are presented in \refsec{sec:results}, followed by concluding remarks and discussion.

Two companion papers describe the deployment of a similar system at Parkes Observatory \citep{PriceBL2017} and the software signal processing, data distribution and management systems for both Green Bank and Parkes \citep{LebofskyBL2017}.

\section{System Architecture}\label{sec:architecture}

\subsection{Green Bank Telescope}

The Robert C. Byrd Green Bank Telescope \citep[GBT;][]{Prestage2009}, is a
100 by 110 meter single dish radio telescope located in Green Bank, West Virginia, USA. The telescope is located within the 34,000~$\mathrm{km}^2$ federally-protected zone National Radio Quiet Zone, in which radio broadcasting is prohibited to minimize Radio Frequency Interference (RFI).  Formerly part of the National Radio Astronomy Observatory (NRAO), as of October 2016 the GBT is now operated by the Green Bank Observatory.

The GBT operates over 0.3--110 GHz, and at any one time it is equipped with a suite of receiver packages that may be selected by the observer; the receivers in use within the BL program are listed in \reftab{tab:receivers}. The GBT has a configurable analog downconversion system, which amplifies, filters and mixes the radio frequency (RF) output from the receiver down to an intermediate frequency (IF) for digitization. 

\subsection{Digital systems}

\begin{figure*}
\begin{center}
\includegraphics[width=1.5\columnwidth]{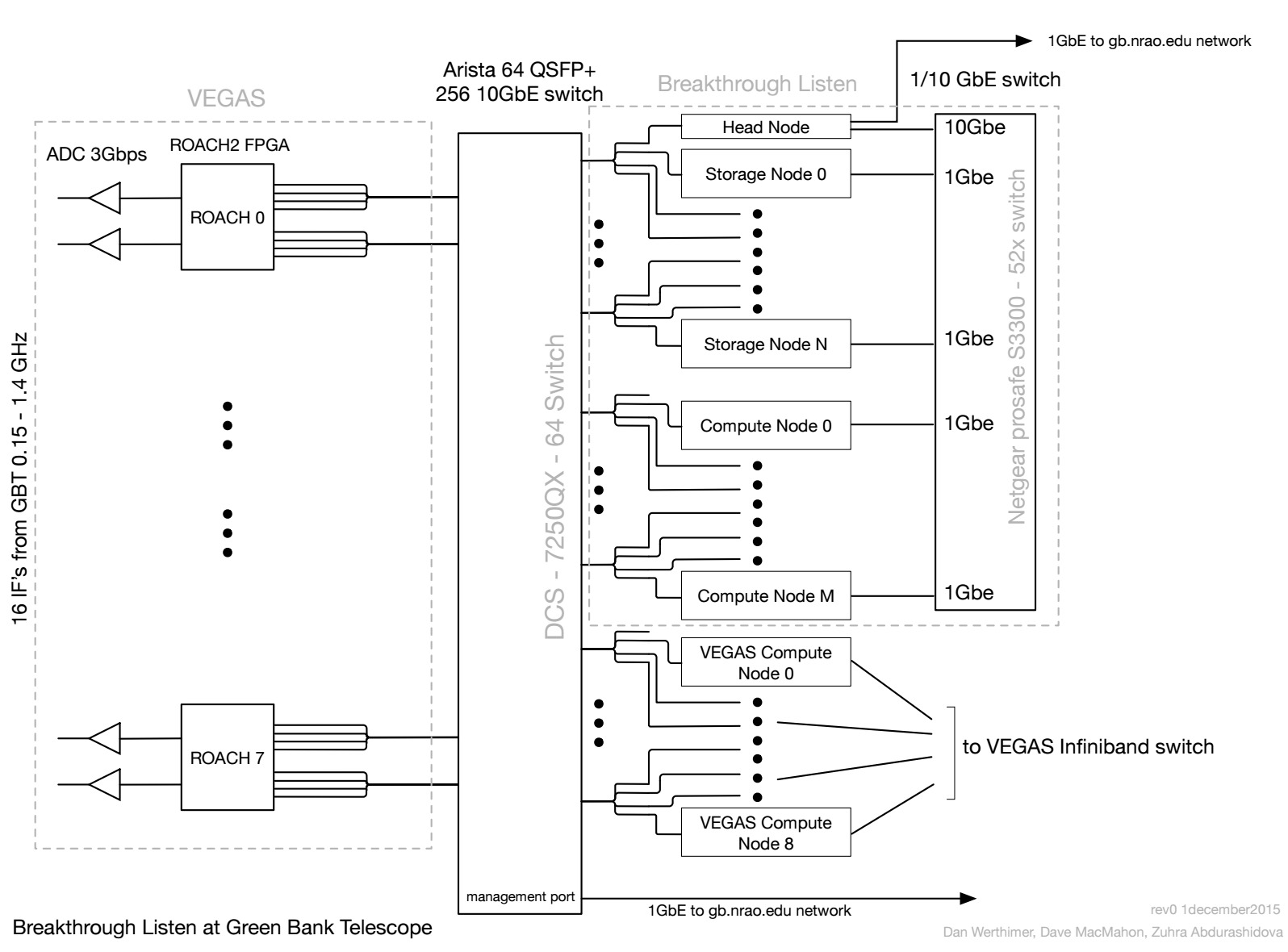}
\protect\caption{\label{fig:architecture} Breakthrough Listen data recorder system architecture}
\end{center}
\end{figure*}

 The BL \dr ~digitizes the signal from the telescope, records the sampled voltages to disk, and performs signal processing tasks for data analysis reduction. The \dr ~employs a heterogeneous architecture in which a field-programmable gate array (FPGA) signal processing `frontend' is connected via high-speed Ethernet to a data capture `backend' consisting of commodity compute servers equipped with graphics processing units (GPUs). A `head node' server is used to provide remote access to the \dr ~system, as well as running system monitor, system control, and metadata collation tasks. A block diagram of the BL \dr ~is shown in \reffig{fig:architecture}.  The basic architecture of recording time domain data products to disk, and performing off-line or pseudo-real-time analysis tasks with them, has been commonly used in searches for extraterrestrial intelligence for several decades \citep{Tarter:1980p1516,Rampadarath:2012fw,2013ApJ...767...94S,EnriquezBL2017}.

Among other outputs, the GBT downconversion system presents eight dual-polarization IF signals to eight ROACH2 signal processing boards developed by the Collaboration for Astronomy Signal Processing and Electronics Research \citep[CASPER,][]{Hickish2016}.  Each ROACH2 board contains two 5 Gsps analog-to-digital converters (ADCs), a Xilinx Virtex6 FPGA, and eight 10 Gbps Ethernet outputs (see \refsec{sec:hardware}). 

\begin{table}
	\caption{GBT receivers in use with (or planned to be used with, in {\em italics}) the BL program \citep{GBTPropGuide2016:2016p6-11}}
	\label{tab:receivers}
	\centering
	\begin{tabular}{l c c c c }
	\hline
	Receiver        & Frequency    & IF bandwidth & Beams & $T_{\rm{sys}}$\footnote{Average across the band, including galactic background and calculated for average weather conditions.} \\
	                & (GHz)          & (GHz) & \#    & (K)       \\
	\hline
	\hline
	L-band          & 1.15-1.73       & 0.58  & 1     & 20\\
	S-band          & 1.73-2.6       & 0.87   & 1    & 25\\
	C-band          & 3.95-8.0       & 3.8   & 1    & 25\\
	X-band          & 8.0-11.6     & 2.4  & 1    & 30\\
	{\em Ku-band}         & 12.0-15.4     & 3.5  & 2    & 30\\
	K-band         & 18.0-27.5     & 4.0  & 7    & 35\\
    {\em Ka-band}        & 26.0-39.5     & 4.0  & 2    & 35\\
	{\em Q-band}         & 38.2-49.8     & 4.0  & 2    & 40\\
	{\em W-band}         & 67.0-93.0     & 4.0  & 2    & 50\\
\hline
	\end{tabular}
\end{table}

The BL \dr ~reuses the ADCs and ROACH2 FPGA boards of the pre-existing Versatile Green Bank Spectrometer \citep[VEGAS,][]{Roshi2011, Prestage2015} as its FPGA frontend; this CASPER-designed hardware is detailed further in \refsec{sec:hardware}. The decision to reuse the VEGAS system decreased hardware and development costs, with the constraint that data transfer must be implemented over high-speed Ethernet. When in use for BL observations, each FPGA is programmed with firmware (see \refsec{sec:firmware}) that digitizes the input signal, applies coarse channelization, and then sends the complex voltage time streams to eight backend compute nodes via 10~Gb Ethernet.

With two ADCs each sampling at 3~Gsps, the Nyquist bandwidth is 1.5~GHz per each polarization. Each backend compute node receives one eighth of this bandwidth as a 187.5~MHz sub-band of critically-sampled complex voltage data for both polarizations on its Ethernet network interface from the FPGA frontend as a stream of UDP (User Datagram Protocol) packets. This results in a 6~Gbps data stream for each compute node.  Each server runs a packet capture code (see \refsec{sec:software}), which parses the UDP header, arranges packets into ascending time order, and writes data into a shared memory ring buffer. The contents of the ring buffer are then written to disk along with telescope metadata that is stored in another shared memory region known as the ``status buffer''.

\subsection{Metadata propagation}

Another process on each compute node, external to the data recording pipeline, periodically pushes the status buffer contents to a \textsc{Redis}\footnote{\url{https://redis.io}} database running on the head node. This process, known as the ``Redis gateway" also subscribes to predefined channels on the Redis database to listen for updates to values stored in the status buffer.  Remote Redis clients can publish updates through these channels to various status buffer fields to orchestrate the operation of the recording pipelines across all compute nodes. Some fields in the status buffer are populated with telescope metadata via a script running on each compute node that communicates with the Green Bank status database.  Having the status buffers from all the compute nodes available in a centralized location is convenient for monitoring the state of the nodes during observations, but the high-throughout demands of the data recording pipeline require that the status buffer be implemented in memory to avoid I/O or networking delays that would be incurred using any sort of file or network based store for the status buffer data. The Redis gateway provides the convenience and utility of a centralized status buffer store while still allowing the high-throughput pipeline to use (fast) local memory for the status buffer.

\subsection{Post-observation data reduction}

During observing sessions, the compute servers record the coarsely channelized data from the FPGA frontend to disk at 8-bits, resulting in a sustained 750~MB/s disk write speed requirement. While fewer bits could be recorded for these intermediate data products, this approach adds complexity to data capture and subsequent analysis. Within the VEGAS FPGA firmware, data grows to 18-bit, before requantization back down to the native 8-bit sampling  of the digitizer. For these intermediate products, storing fewer than 8 bits only servers to complicate further data reduction and analysis. Namely, the smallest integer datatype in C and Python is 8-bit; smaller representations must be promoted to 8-bit for any manipulation (at a non-zero computational cost). Additionally, the dynamic range is limited significantly, and careful (and dynamic) equalization of the passband is required; quantization efficiency is lower so signal-to-noise increases; and non-linear quantization gain must also be taken into account. Between observations, the recorded 8-bit complex voltages are converted into reduced data products using a GPU-accelerated code (\refsec{sec:software}) reducing the data to approximately 2\% of its original size. The reduced data products, and a selection of voltage data of interest, is sent to storage servers for archiving and the disks on the compute nodes are cleared in preparation for the next observation. Work is underway to run real-time data analysis routines during observations.

\section{Deployment Strategy}\label{sec:deployment}

For rapid installation and to allow iterative improvements, we employed a staged deployment strategy for the \dr ~compute and storage servers. The overall bandwidth of the \dr ~system is determined by the number of compute servers attached to VEGAS via the Ethernet network; each server receives and processes a discrete 187.5~MHz. A first-stage system was deployed in July 2015, consisting of one compute nodes and one storage server. In January 2016 the system was expanded to eight compute nodes capable of recording 1.5 GHz of bandwidth.  Throughout 2016, the system was expanded in several stages.  As of April 2017, the system consists of 32 compute nodes, with an instantaneous recording capability of 6 GHz, and 4 storage nodes.  Later in 2017, the system will be doubled again to enable recording of a 12 GHz Nyquist bandwidth, though analog filtering limits the usable bandwidth to approximately 10 GHz.

As detailed in \cite{Isaacson2017}, early BL observations consist of a survey of 1709 nearby stars with the L, S, C and X-band receivers. Early observations focused on use of the L and S-band receivers, as their intrinsically smaller bandwidth allowed for the full bandwidth to be recorded with the first-phase BL \dr.

\section{Digital Hardware}\label{sec:hardware}

The BL \dr ~is comprised by off-the-shelf compute and storage servers, Ethernet networking hardware, and FPGA boards developed by CASPER.
This hardware is introduced further below.

\begin{figure}
\includegraphics[width=0.85\columnwidth]{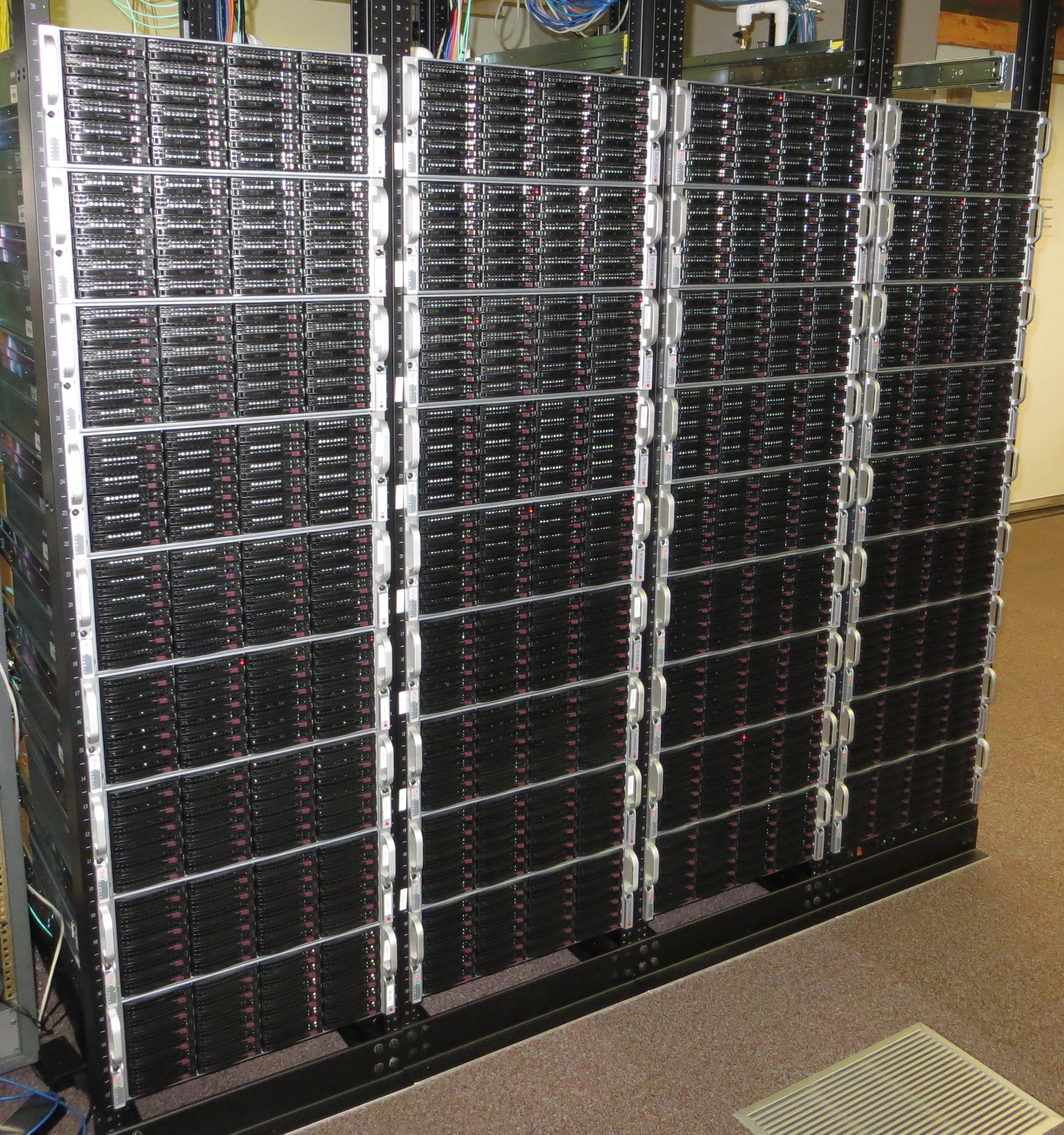}
\caption{\label{fig:bl-rack-front} BL compute nodes, front.
}
\end{figure}

\begin{figure}
\includegraphics[width=0.83\columnwidth]{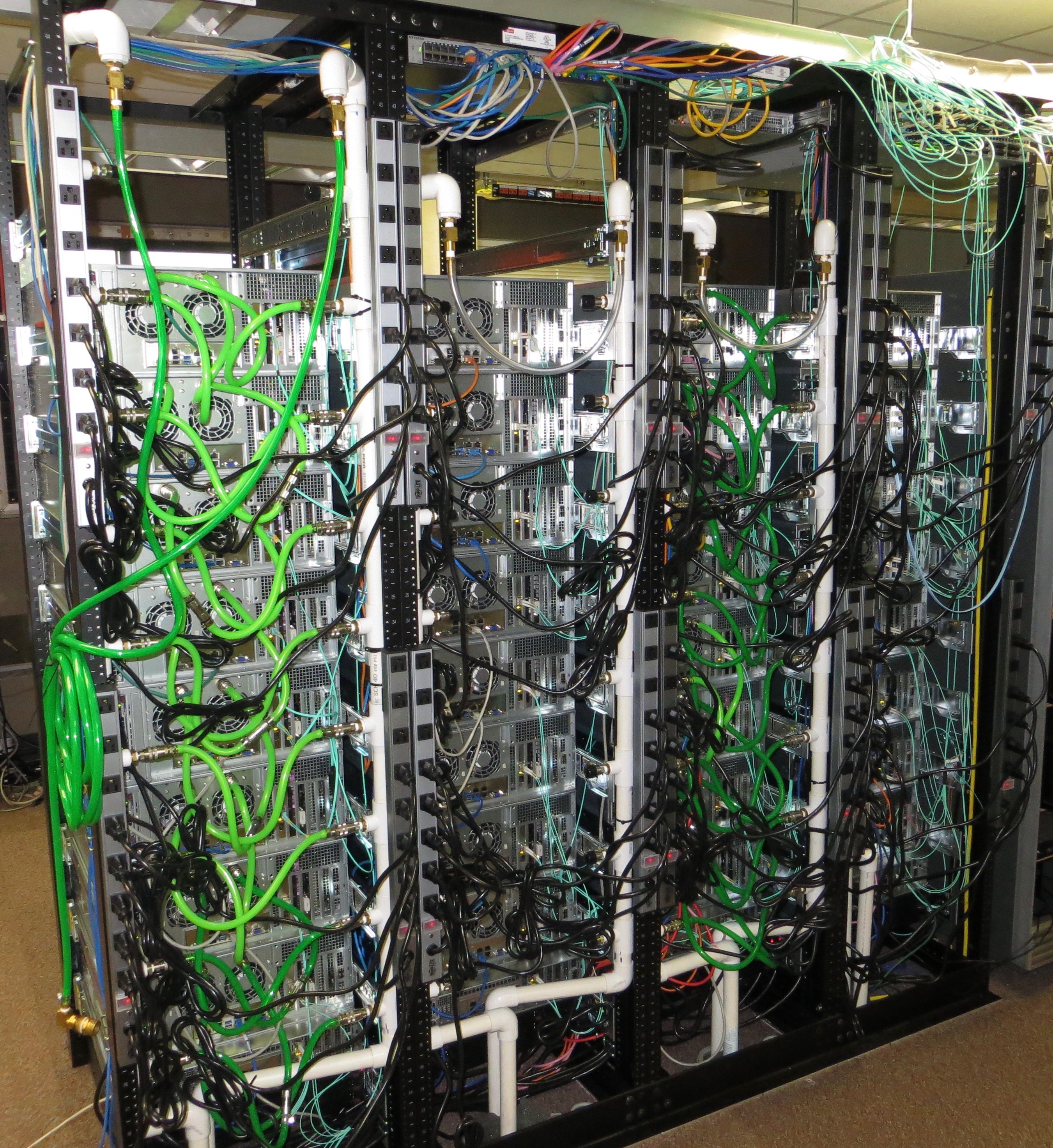}
\caption{\label{fig:bl-rack-back} BL compute nodes, back.
}
\end{figure}

\subsection{ADC card}

The VEGAS front-end uses a set of analog-to-digital converter (ADC) cards to digitize the signal presented by the telescope's analog chain. To digitize each dual-polarization IF signal, two of the CASPER ADC5G daughter cards \citep{Patel2014} are connected to a ROACH2 FPGA processing board. The ADC5G is an 8-bit, 5~Gsps ADC card based upon the e2V EV8AQ160 chip. Because the 1.5 GHz wide IF signal presented to the ADCs occupies 0--1.5 GHz, BL operates these ADCs at the Nyquist rate of 3~Gsps.

The ADC5G was selected for its wide bandwidth and well characterized performance in radio astronomy applications \citep{Patel2014}.  This ADC has been used in several other radio astronomy projects, such as the SWARM correlator at the Submillimeter Array \citep{doi:10.1142/S2251171716410063} and ROACH2 digital backend \citep[R2DBE, ][]{2015PASP..127.1226V}. 

\subsection{ROACH2 FPGA board}

The CASPER ROACH2 -- Reconfigurable Open-Architecture Compute Hardware version 2 -- is an FPGA-based signal processing platform \citep{Hickish2016}. The ROACH2 features a Xilinx Virtex-6 SX475T FPGA. Each ROACH2 is equipped with two ADC5G daughter cards, and eight 10~GbE SFP+ Ethernet interfaces. 

Monitor and control of the ROACH2 is conducted via an on-board PowerPC that runs a lightweight variant of Debian Linux operating system provided by CASPER. The PowerPC allows for the FPGA to be reprogrammed as and when required. After programming, control registers of the FPGA runtime design are made available on the PowerPC via a memory-mapped interface. The ROACH2 boards can be controlled remotely via the use of the Karoo Array Telescope Control Protocol (KATCP\footnote{\url{https://casper.berkeley.edu/wiki/KATCP}}), which provides a networked interface to the memory mapped control registers and memory regions.

\subsection{PPS and clock synthesizer}

Each ROACH2 board requires a reference `clock' frequency standard to be provided to drive the ADC; the FPGA clock is also derived from the ADC clock signal. These clock signals are generated by Valon\footnote{\url{http://www.valontechnology.com}} 5007 synthesizers, one per ROACH2 board.  For each ROACH2, the clock signal is distributed to the two ADCs via a power divider network. The Valon synthesizers are locked to a 10~MHz frequency reference signal from an on-site hydrogen maser.

In addition to the reference clock, a pulse-per-second (PPS) signal derived from the Global Positioning System (GPS) is also distributed to each board. By doing so, boards may be armed to start data capture on the positive edge of the PPS.  Even so, a number of factors conspire to make calibration between boards and/or to absolute time difficult. In addition to the usual cable propagation differences between the sky and 1 PPS signals, the ADC samples are de-multiplexed by two factors of four which gives rise to an inherent delay ambiguity of $\pm 8$ ADC samples. Analysis techniques for which this presents a challenge (e.g. interferometry, pulsar timing) must perform on-sky calibration to resolve these ambiguities, which are stable until the next arming of the 1 PPS signal.

\subsection{Ethernet interconnect}

All ROACH2 boards and compute nodes (see below) are connected together via a 10~Gb Ethernet network. Full-crossbar connectivity is provided by an Arista 7250QX-64 switch, which has a total of 64 QSFP+ (40~GbE) ports. This switch is shared with other signal processing backends that form the VEGAS instrument. Each QSFP+ port on the switch is converted to four 10 Gbps SFP+ ports via industry-standard `break out' cables.

Remote access to the BL \dr ~system is provided by a 1~Gb connection to the head node; a secondary 1~Gb Ethernet interface connects the head node to an internal network that provides monitor and control for the FPGA and compute nodes. For high-speed off-site data transfer, a 10~Gb connection on the 7250QX-64 switch to the NRAO network may be used.

The VEGAS FPGA frontend produces a packetized stream of UDP packets with data payloads as large as 8192 bytes. As such, to allow the transmission of `jumbo frames', the MTU on all interfaces is set to 9000 bytes.

\subsection{Compute servers}

\begin{table}
	\caption{BL \dr ~compute node configuration.}
	\label{tab:compute}
	\centering
	\begin{tabular}{l r}
	\hline
	Chassis         & Supermicro 4U 6048R-E1CR24N \\
	Motherboard     & Supermicro X10DRi-T4+ \\
	CPU             & Intel Xeon E5-2620 v4 \\
	GPU             & NVIDIA GTX 1080 \\
	NIC             & Mellanox MCX312A-XCBT \\
	Memory          & 64 GB DDR4   \\
	Hard drive      & 24 Toshiba X300 5TB \\
	\hline
	\end{tabular}
\end{table}

\begin{table}
	\caption{BL \dr ~storage node configuration.}
	\label{tab:storage}
	\centering
	\begin{tabular}{l r}
	\hline
	Chassis         & Supermicro 4U 6048R-E1CR36N \\
	Motherboard     & Supermicro X10DRi-T4+ \\
	CPU             & Intel Xeon E5-2620 v4 \\
	NIC             & Mellanox MCX312A-XCBT \\
	Memory          & 64 GB DDR4   \\
	Hard drive      & 36 Toshiba HDWE160 6TB \\
	\hline
	\end{tabular}
\end{table}

The overall bandwidth of the BL \dr ~system is determined by the number of compute servers installed, with each server processing 187.5~MHz of bandwidth. We selected Supermicro brand systems, each housed in a 4 rack unit (RU) chassis. Due to the staged deployment, there are some minor differences between server configurations. Notably, systems installed in the first phase have NVIDIA Titan X Maxwell GPUs, as the newer NVIDIA GTX 1080 Pascal cards were not yet released.

\subsection{Storage servers}

The storage servers are similar in configuration to the compute servers, but use a different Supermicro chassis (6048R-E1CR36N) which supports up to 36 hard drives.  As no signal processing is performed on the storage servers, no discrete GPUs are installed.

Each storage server contains 36 hard drives, each with 6 TB capacity. These are configured as three 11-disk RAID 6 arrays plus 3 global hot swap spares. The RAID volumes are formatted with XFS filesystems. This results in 50 TB of usable storage capacity per RAID volume or 150 TB of usable storage capacity per storage node.

A ratio of one storage server for every eight compute servers is used. For long-term archiving and public access, data from the storage servers will be transferred to an open data archive that is currently under development.

The current storage server configuration is summarized in \reftab{tab:storage}.

\subsection{Power and cooling}

The BL \dr ~system is installed in an RFI-shielded building approximately 1 mile from the GBT. Heat waste from the BL \dr ~is removed via both air conditioning and closed-loop direct-to-chip water cooling (\reffig{fig:bl-rack-back}). The water cooled compute nodes have passive water blocks on the two CPU chips. The GPUs are purchased with OEM-installed water blocks, rather than retrofitting third party water blocks onto air cooled GPUs.

The total power usage envelope of the \dr ~is 48.4~kW; a breakdown of the power budget is given in \reftab{tab:power}.

\begin{table}
	\caption{BL \dr ~power budget, not including cooling.}
	\label{tab:power}
	\centering
	\begin{tabular}{l c c r}
	\hline
	Item           & Quantity  & Power/unit & Total \\
	               &     -     & (W)        & (kW)  \\
	\hline
	\hline
	VEGAS (total)  &  8        &  200      &  1.6  \\
	Head node      &  1        &  400      &  0.4  \\
	Compute nodes  & 64        &  600      & 38.4  \\
	Storage nodes  &  8        &  500      &  4.0  \\
	Storage nodes  &  8        &  500      &  4.0  \\
	\hline
	               &           &           & \textbf{48.4}
	\end{tabular}
\end{table}

\section{FPGA Firmware}\label{sec:firmware}

The FPGA firmware used for the \dr ~is the same as is used for 512-channel VEGAS Pulsar Mode. This design, developed by Green Bank Observatory staff using the CASPER toolflow, runs on the CASPER ROACH2 FPGA boards. Each ROACH2 board digitizes two IF signals which are typically configured to be both polarizations of a 1.5 GHz bandwidth IF signal. The current system can fully ingest the output from up to four ROACH2s, for an aggregate Nyquist bandwidth of 6~GHz.  The digitized signals are divided into 512 frequency channels using an 8x overlapped Polyphase Filter Bank (PFB) from the CASPER DSP library.  Each (now complex) signal is requantized to 8 bits real and 8 bits imaginary. The 512 channels are distributed via UDP packets through the ROACH2's eight 10 GbE interfaces through the 10 GbE switch to eight different Breakthrough Listen compute nodes, each receiving a different 64 channel subset of the data.  When the same FPGA design is used for VEGAS pulsar mode, the destination IP addresses in the ROACH2 are configured to direct the UDP packets to the VEGAS backend systems instead.  Sharing the same FPGA hardware and runtime design greatly sped the development and deployment of the BL \dr.

\section{Software}\label{sec:software}

\subsection{HASHPIPE data capture}

The \textsc{hashpipe}\footnote{\url{https://github.com/david-macmahon/hashpipe}} software package is used to capture UDP packets, arrange them in a meaningful way, and write the data to  disk. \textsc{hashpipe} is written in C and implements an in-memory ring buffer through which data can be shared among different threads of one or more processes.  \textsc{hashpipe} is an evolution of the VEGAS HPC software pipeline that is itself derived from the GUPPI data acquisition pipeline \citep{2008SPIE.7019E..1DD}.  \textsc{hashpipe} is a framework that handles the system level functionality of setting up shared memory segments and semaphores.  Applications written for \textsc{hashpipe} are created as shared library ``plug-ins".  \textsc{hashpipe} offers features not found in its ancestral predecessors such as the ability to run multiple instances on a single host and the ability to construct multi-threaded pipelines and assign processor affinities on the command line.

In our implementation, a copy of our \textsc{hashpipe}-based pipeline is launched on each compute server. During an observation, the pipeline on each server captures UDP packets and parses their headers to determine how to copy the packet contents into the ring buffer.  The packets from the FPGA all have timestamps so they can be reordered into the proper time sequence if they happen to arrive out of order from the network.  Once a contiguous block of 128 MiB of data has been written to the ring buffer, the block is handed to the next thread for writing to disk. 

Recorded data are written to disk in the GUPPI raw format \citep{Ford:2010p8485}. This consists a simple plain text header, loosely based on the FITS format \cite{Pence2010}, followed by a block of binary data. This header plus data block format is repeated a total of 128 times at which time the data file is closed and another one is created. Each data file is slightly larger than 16 GiB and contains approximately 23 seconds of complex dual polarization sky voltage data (except the last which is shorter).  This data format, and all subsequent data reduction  is described in \cite{LebofskyBL2017}.

\begin{figure}
\includegraphics[width=1\columnwidth]{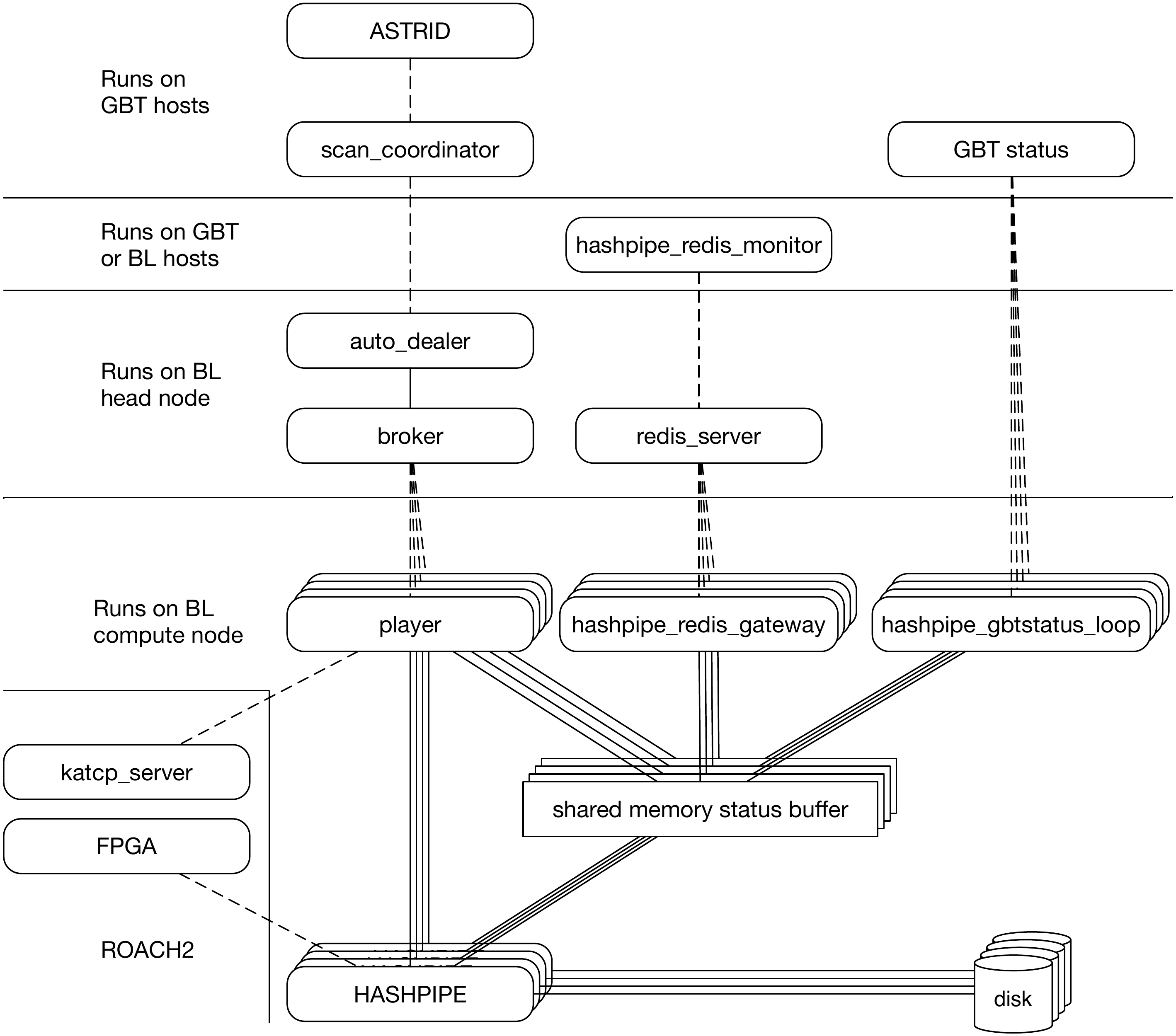}
\protect\caption{\label{fig:sw-arch} Breakthrough Listen Green Bank Data Recorder software architecture}
\end{figure}

%\subsection{Data reduction and analysis}
%Suggest high level description here with technical details in Lebofsky et al.

\subsection{Systems monitoring}

% \emph{Assigned to Matt ---}
% Mention ganglia, astrid and observing scripts here.
To ensure system health and maximum efficiency during observations (and between observations, when data reduction and analysis takes place), we use various monitoring tools. 

Many such open source monitoring systems exist, but given its ease and flexibility we settled upon employing \textsc{Ganglia}\footnote{\url{http://ganglia.sourceforge.net/}} for checking various general statistics (system load, network, disk storage, uptime). \textsc{Ganglia} also allows for very easy implementation of gathering more project specific statistics, such as GPU temperatures, average system fan speeds, or whether critical parts of the analysis pipeline are running successfully.

In addition, we have developed our own set of monitoring scripts, run as cron jobs or daemons, that check various pipeline or systems pulses and act accordingly. Actions include: shutting systems down immediately if the temperatures of critical components rise above threshold levels.  

The GBT employs a telescope-control system and known as \textsc{Astrid} with which observations are managed and controlled. During observations, our \textsc{Astrid} scripts are written with extra logic to avoid repeating targets in case of a failure/restart. We are currently tightening the integration between \textsc{Astrid} and the Breakthrough Listen backend system to prevent telescope time and data loss due to lack of error propagation from BL \dr ~compute nodes back to \textsc{Astrid}.  

\section{Telescope integration}
\label{sec:integration}

Integration with the GBT control system is crucial to the operation of the \dr.  Observational metadata from the telescope control system (pointing location, frequency tuning, timestamps, etc.) are necessary for the recorded sky data to be scientifically useful.  Furthermore, knowing when to start and stop recording for each observed source is necessary to conserve disk space and ensure that datasets are more manageable in size.  This integration with the facility's telescope control system (TCS) was accomplished by taking advantage of different access paths into the TCS.  Listening to messages published by the GBT's ``ScanCoordinator'' process provides the necessary information regarding to start and stop times for recording data from each observed source.  Simultaneously, periodic queries to the GBT status database provide relevant real-time details that are recorded synchronously with the voltage data from the sky.

Green Bank Observatory provided Python code that listens to ScanCoordinator and starts/stops the data recording during scans.  The component of this software that listens to ScanCoordinator is referred to as ``the dealer'' and the component that starts/stops the recording on a compute node is referred to as ``a player'' \footnote{a continuance of the (Las) VEGAS theme}.  The ScanCoordinator communicates with various clients, including the dealer, using the ZeroMQ\footnote{http://zeromq.org} messaging protocol.

Other Python code communicates with the GBT's ``IFManager" to determine the frequencies being observed.  This information is used to compute frequency related metadata that get stored in each compute nodes shared memory status buffer that ultimately gets saved as the header blocks in the recorded data files.  These values are updated whenever the telescope's downconversion system is configured, which is currently done once per session for a given receiver since the tuning is not changed during observations (e.g. no Doppler tracking is performed by the local oscillators).

Metadata that changes during a scan (e.g. azimuth and zenith angle) are queried periodically ($\sim$1~Hz) from the GBT's status database.  These values are also stored in the shared memory status buffer and ultimately included in data file headers.

\section{Performance tuning}
\label{sec:performance}

Here we discuss some of the tuning necessary to ensure the compute nodes can sustain the necessary network-to-disk throughput. A prerequisite for performance tuning is understanding the architecture of the hardware and software upon and with which the the system is built. The Intel Xeon multi-socket main boards used in this project have a non-uniform memory access (NUMA) architecture, meaning that memory regions and peripherals are more closely associated with a specific CPU socket than with others. This non-uniformity also include the PCIe slots and their associated interrupts. A CPU and its closely associated resources form what is known as a NUMA node. Performance is best when data do not have to migrate between NUMA nodes, but this does not happen automatically. These systems are designed to make their non-uniformity fully transparent, but when pushing performance the mechanisms that provide this transparency become a bottleneck. Fortunately, modern operating systems provide ways to exercise control over the system to optimize performance in a NUMA-aware manner. 

For the BL \dr, the main operations are: 
\begin{itemize}
	\item Capture Ethernet packets from the network.
	\item Arrange the packet data in memory to recreate original timestream.
	\item Write the data from volatile memory to disk.
\end{itemize}

Optimizations can be applied to all these operations.  They are presented here in the same order that is seen by data flowing through the system.  

The first optimization is to change how the network interface card (NIC) interrupts the CPU.  Instead of interrupting on every received packet, most modern NICs can be configured to interrupt the CPU every $N$ packets or every $M$ bytes of data.  This is a process known as  ``interrupt coalescing'' and can be configured on Linux using the \texttt{ethtool} utility.  Interrupt coalescing can increase the apparent latency of the NIC, but the BL \dr ~just needs to maintain sustained throughput; latency is not a relevant issue.

When the NIC does interrupt the CPU, it is desirable to ensure that the interrupt is handled by a CPU core running on the same NUMA node as the NIC.  On Linux, the distribution of interrupts to CPU cores can be seen in the \texttt{/proc/interrupts} file.  Each interrupt can be limited to a specific subset of CPU cores by writing an to the \texttt{/proc/irq/<N>/smp\_affinity} file.  We found the best performance when directing the NIC interrupt requests (IRQs) to the same processor that runs the thread capturing data packets (see below).

Using the portable \texttt{recv()} system call to receive UDP packets requires that the packets pass through the kernel's IP stack.  While this works, we found that the time spent in the \texttt{recv()} call varied far more than we expected; sometimes more than we could tolerate.  Several proprietary solutions to this exist (Intel's Data Plane Development Kit\footnote{\url{https://www.dpdk.org}} as well as NIC vendor accelerator libraries), but they all involve acquiring and running proprietary code from Intel or the NIC's vendor.  Linux provides a middle ground option known as ``memory mapped packet sockets''.  Packet sockets operate a very low level in the kernel's network stack, avoiding much of its overhead.  When using memory mapped packet sockets, a memory buffer is pre-allocated in kernel space and mapped into the user process's address space via \texttt{mmap}.  Incoming packets are copied into a part of this buffer and a bit is set to indicate that that part of the buffer now belongs to the application.  The application much clear that bit before the corresponding part is handed back to the kernel.  \textsc{Hashpipe} provides high level wrappers around the setup and handing of packet sockets.  More info on using memory mapped packet sockets can be found in the \texttt{networking/packet\_mmap.txt} file of the Linux kernel \texttt{Documentation} directory.

One detail to be aware of when using packet sockets to receive UDP packets is that the UDP packets still traverse the kernel's IP stack.  Since it is unlikely that any other process will be listening for these packets with \texttt{recv()}, the kernel will send ``destination unreachable'' ICMP packets back to the sender, resulting in more unnecessary 10 GbE traffic that could contribute to packet loss.  The solution to this potential problem is to use the \texttt{iptables} firewall to drop the incoming UDP packets (packet sockets will see the packets before the firewall drops them).

Once the UDP packets are delivered to the user process, the next step is to copy the UDP payload from the packet socket buffer to the shared memory data block to build up the 128 MiB blocks of data.  As mentioned above, we found best performance when binding (aka pinning) the network thread to the same CPU core as is used to handle the NIC interrupts.  \textsc{Hashpipe} makes it easy to specify the processor affinity of each thread in the pipeline.  The data samples in the UDP packets needs to be partially transposed because of data ordering differences between the FPGA packet format and the GUPPI RAW data format.  When performing this operation, we obtained better performance by writing the transpose loop with an eye toward minimizing cache thrashing.  If needed, ``non-temporal'' write instructions, part of the Intel instruction set, can be utilized to prevent cache pollution altogether.

As mentioned elsewhere, the data are saved to files on RAID5 volumes formatted with XFS file systems.  By benchmarking the effects on performance of various XFS options, we were able to determine that keeping the XFS file system \emph{un}aware of the RAID5 volume's stripe size gave the best performance.  We also found that the default buffering of the file I/O on Linux resulted is very bursty loads on the disks that could starve other resources for system time.  Using Direct I/O (i.e. passing \texttt{O\_DIRECT} to the \texttt{open()} system call) avoids the file buffering and caching that normally takes place and data are written directly to disk with each call to \texttt{write()}.  This resulted in a far more constant and consistent load on the disks and the disk writing CPU.  Direct I/O imposes a constraint that data must be written in multiples of 512 bytes.  For our data blocks this was fine since they are inherently a multiple of 512, but for writing the headers (which consist of 80 byte records) it generally requires adding padding bytes to the end of the header to round it out to the next multiple of 512 bytes.  To inform future readers of the file that such padding exists, we add a \texttt{DIRECTIO=1} header entry.  Readers must examine the header for this setting and, if found and non-zero, seek forward past the padding to get to the start of the data block.

\subsection{Ethernet verification tests}

To confirm the functionality of the Ethernet interfaces on the servers, we performed port-to-port tests between server pairs, routed via the switch. We used the Linux utility \texttt{iperf} to generate and receive UDP packets, and confirmed $>$9~Gbps throughput using jumbo frames ($>$8192 B). A simple \textsc{Hashpipe}-based pipeline was also used to confirm data could be captured at the requisite 6~Gbps without packet loss.

The Arista 7250QX-64 switch is rated to handle up to 5~Tbps data throughput; the requirement during observations is 320~Gbps (10~GHz bandwidth at 8-bit, dual inputs), a small fraction of the overall capability. For initial functionality testing, we implemented ROACH-2 firmware  to generate 8 streams of `dummy' 10 GbE traffic at line rate. Each of the ROACH-2 Ethernet interfaces sends dummy data to another interface via the switch, and a simple checksum test is run by the firmware to confirm data validity. We used this firmware in the laboratory to verify that the switch functions without packet loss at speeds of up to 9.8~Gbps/port.

Ethernet performance can be monitored in several ways.  On the macroscopic scale, the Ethernet switch can be queried to determine the ingress and egress data rates for the ROACH2 and compute node ports.  Any ports that are transmitting or receiving more or fewer packets per second may be indicative of a problem.  On the receive side, the packet sequence numbers in the first 8 bytes of the UDP payload can be tracked to ensure that all packets are received.  Any gaps indicate missed packets.  Because the packet sequence numbers are used to arrange the packets in memory, the packet sequence tracking can be performed simply by counting the packtets received for each data block.  Gaps can be detected as fewer packets per data block.  With the system performance tuned as described in section \ref{sec:performance}, packet loss is negligible to non-existant.

\subsection{Disk write tests}

Since recording 6 Gbps/node requires fast disk write speeds, we had to tune our systems to ensure this was possible. Tuning involves different RAID types, sizes, and striping, as well as different filesystems. Eventually we settled on each system having two 11-drive RAID5s (which allowed for resilience, enough speed, and still two global hot spares in each system). On these RAIDs we installed XFS filesystems (after finding ext4's sustained throughput to be too erratic).

We confirmed disk speeds using basic ``dd" commands, i.e.: 

\noindent
\begin{verbatim}
dd if=/dev/zero bs=1M  \
of=/datax/test.out  \
oflag=direct,sync
\end{verbatim}
This brought to light other tunable parameters which affect performance (see \refsec{sec:performance}). For example, we found that XFS stripe sizes somehow interfere with the RAID's striping. We got the best performance by using the \texttt{noalign} option when creating the XFS filesystems. We find a maximum disk write speed of $\sim$16~Gbps per RAID array, leaving ample headroom for capturing the required 6~Gbps per node.

\section{System verification}
\label{sec:results}

For system verification, we conducted observations of known astrophysical and artificial sources. The VEGAS frontend has been tested extensively and is in use as a facility instrument; as such, we focus on \dr-specific verification in this section.

\subsection{Tone injection}

The GBT allows for a test tone (i.e. sine wave of configurable amplitude and frequency) to be injected at an early stage of the analog receiver system. We injected test tones to verify that frequency metadata was correctly propagated from the IFManager through to the \dr. By this approach, we confirmed our channel frequencies matched the output of the signal generator to better than our channel resolution (3~Hz).

\subsection{ADC histograms}

\begin{figure}
\includegraphics[width=1.0\columnwidth]{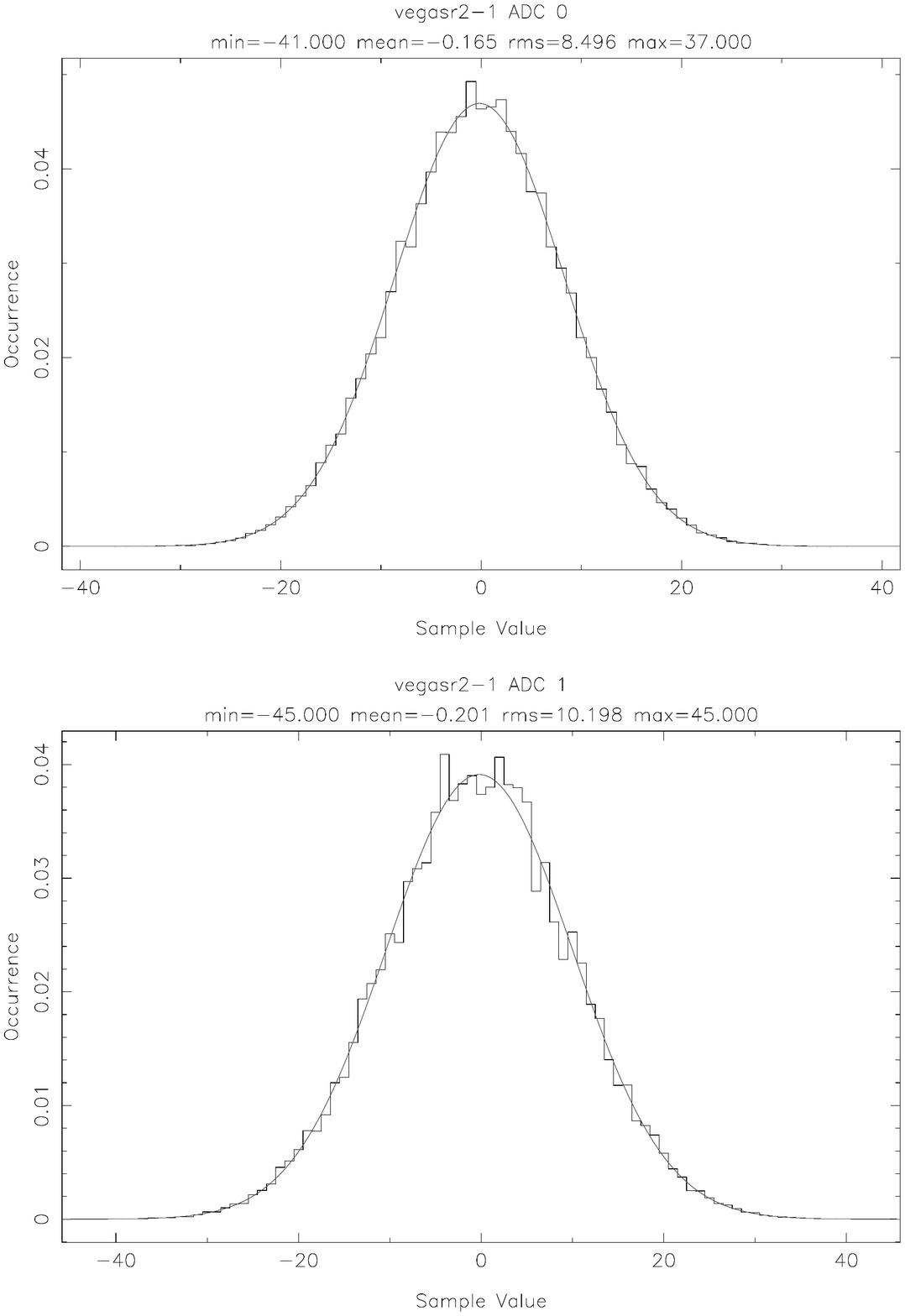}
\caption{\label{fig:adc-histo} An ADC histogram for a typical BL \dr ~observation at L-band, taken after balancing power levels via ASTRID.
}
\end{figure}

 The GBT observing tool, ASTRID, provides a ``Balance()" command that adjusts the power levels throughout the IF system to avoid saturation and and other non-linearities along the components comprising the signal path.  In the absence of strong interference, the probability distribution for a noise-dominated astronomical signal is close to Gaussian, and an optimal input RMS power level to the ADC may be computed \citep{Thompson2007}. We target an RMS level of 12.8 ADC counts (18 mVpp), which gives high quantization efficiency (better than 99.94\%) while leaving headroom for impulsive interference. A histogram of  ADC samples showing a characteristic Gaussian shape is shown in \reffig{fig:adc-histo}.

\subsection{Pulsar observations}\label{sub:pulsar}
\begin{figure}
\begin{center}
\centering
\includegraphics[scale=0.35,angle=-90]{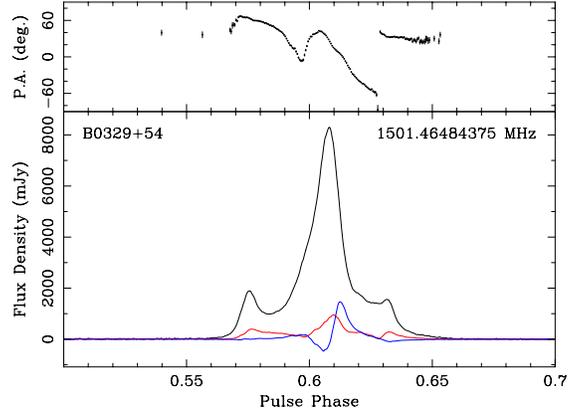}
\protect\caption{\label{calib_prof} Calibrated folded profile of PSR B0329+54 observed using the BL backend 
at L-band. The bottom panel shows folded profiles for total power (black solid line), linear polarization (red solid line) and 
circular polarization (blue solid line). The top panel shows the polarization position angle (P.A.) as a function of pulse phase.}
\end{center}
\end{figure}

\begin{figure}
\begin{center}
\centering
\includegraphics[scale=0.35,angle=-90]{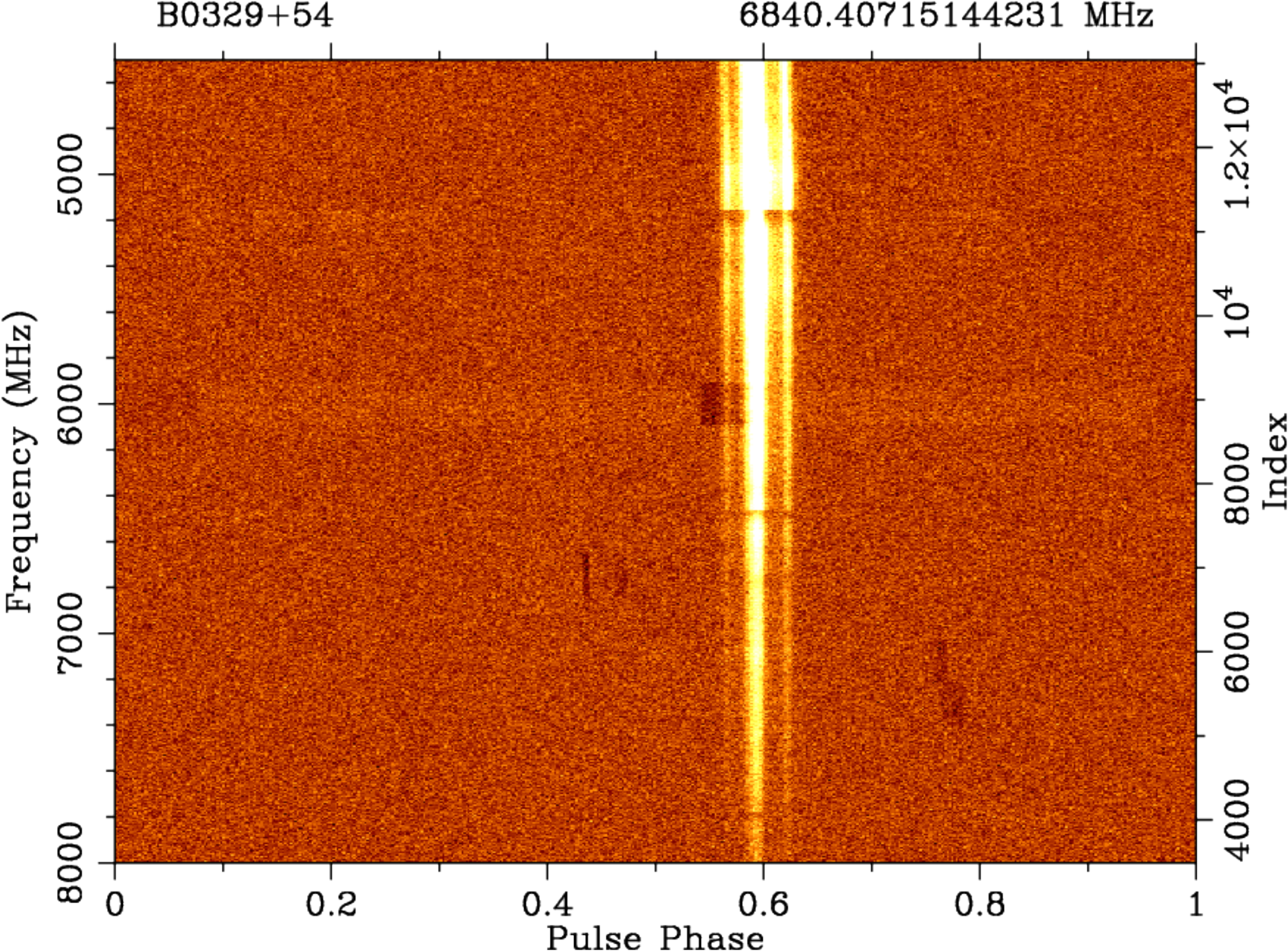}
\protect\caption{\label{cohdedisp_0329} Coherently dedispersed folded profile across 3.2 GHz of PSR B0329+54, observed using the BL DR at the C-band. The plot shows folded intensities at each frequency as a function of pulse phase. Notice the gradual decline in the pulse intensity towards higher frequencies due to the steep negative spectral index for the pulsar.}
\end{center}
\end{figure}

In order to commission the BL \dr ~for high precision pulsar science, we carried out observations of several pulsars. We also carried out a detail polarization and flux measurement of PSR B0329+54, the brightest pulsar in the northern hemisphere, in order to scrutinize our backend's behavior.  We used a switching noise diode \citep{GBTObsGuide2017} along with observations of flux calibrator 3C138.  A fully calibrated folded profile along with the linear and circular polarization profiles are shown in Figure \ref{calib_prof}.  The previously known flux on this pulsar was around 203$\pm$57~mJy \citep{lyl+95} which matches with our measured flux of around  178$\pm$11~mJy.  We also found consistent polarization-angle (P.A.) swing compared to the earlier reported behavior by \cite{gl98}, which demonstrate the polarization capabilities of our instrument. 
%Recently, we expanded the bandwidth coverage of our backend in order to cover a larger part of the available receivers' IF bandwidths at higher frequency bands (see Table \ref{tab:receivers}). 
Figure \ref{cohdedisp_0329} shows a typical observation of PSR B0329$+$54, carried out at C-band across 3.2 GHz of bandwidth. 
%We also carried out observations with our entire backend capabilities of 4.8 GHz and coherently dedisperse 
%the compute nodes. 
%{\bf Things to put : Polarization Cal profile. Single pulse text. Spliced data across 4.8 GHz bandwidth of 
%recent C-band observations.}
%IMAGES: show single pulse, folded pulse profile.

%\begin{figure}
%\begin{center}
%\centering
%\includegraphics[scale=0.35,angle=0]{Pulse2.pdf}
%\protect\caption{\label{fig:pulsar} A pulse.}
%\end{center}
%\end{figure}

%\begin{figure}
%\includegraphics[width=1\columnwidth]{figures/pulsar}
%\protect\caption{\label{fig:pulsar} A pulsar.
%}
%\end{figure}

\subsection{Voyager 1}

\begin{figure}
\includegraphics[width=1\columnwidth]{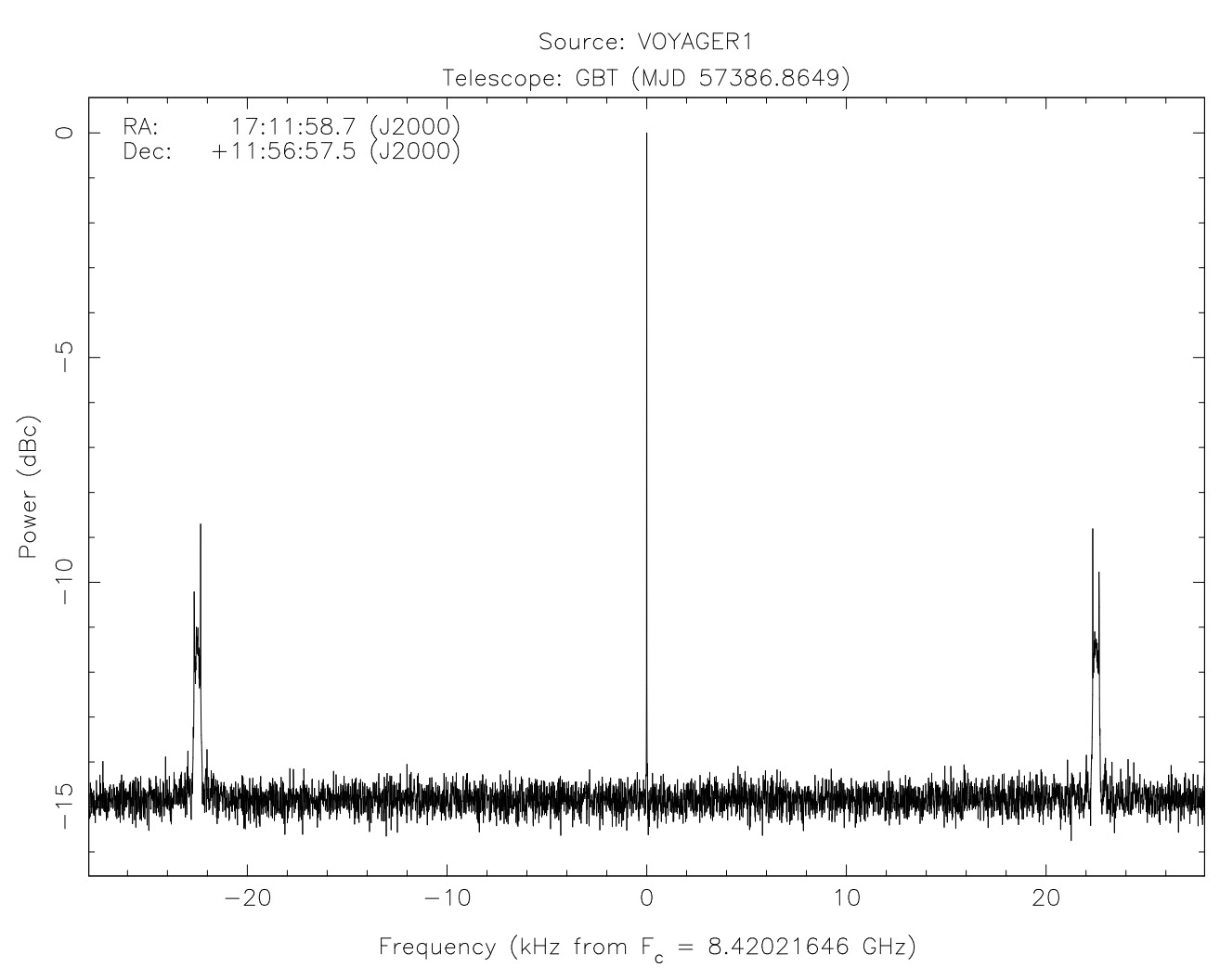}
\protect\caption{\label{fig:voyager} Power spectrum for a 2 minute diagnostic observation of the Voyager I spacecraft conducted with the Breakthrough Listen \dr.   Uncalibrated total power (Stokes I) is plotted vs. topocentric frequency, although this transmitter is intrinsically circularly polarized.  Both the center carrier and sidebands (22.5 kHz offset) are visible.
}
\end{figure}

The Voyager 1 spacecraft was launched September 5, 1977.  As of March 2017, it is over 138 AU from Earth, making it the most distant human-made object.  Voyager 1 is still transmitting a radio telemetry signal, almost 40 years after launch.  As part of commissioning the BL \dr ~system on GBT, at UTC 20:45 2015-12-30, we observed the computed ephemeris of Voyager 1 using the X-band receiver. The telemetry signal from Voyager was detected at high signal-to-noise (\reffig{fig:voyager}), clearly showing the carrier and sidebands.  Such detections  of spaceborne human technology are important tests of the BL \dr, in that they are the closest analogues to the signals being sought in radio searches for extraterrestrial technologies. 

\section{Discussion}

\subsection{Comparisons With Other Instruments}
Table \ref{tab:compare} compares  the current BL \dr ~with other radio astronomy instruments performing high speed coherent voltage recording to disk.  To our knowledge, the BL \dr ~has the highest maximum data recording rate  of any radio astronomy instrument.

\begin{table*}
	\caption{Specifications for selected radio astronomy instruments performing baseband data recording, including the BL \dr.}
	\label{tab:compare}
	\centering
	\begin{tabular}{l c c c c}
	\hline
    Instrument & Bits & Bandwidth & Beams/IFs & Total Data Rate \\  
     & & (MHz) & & (Gbps) \\
    \hline
    \hline
    Breakthrough Listen (GBT, current) & 8 & 1500-6000 & 2-8 & 192 \\
	SETI@home Data Recorder (Arecibo)\footnote{\cite{2009ASPC..420..431K}} & 1 & 2.5 & 26 & 0.13 \\
	LOFAR\footnote{\cite{cobalturl}} & NA & NA & NA & 40\footnote{LOFAR is extremely flexible and operates in a variety of modes, but is data rate limited.} \\
	GUPPI/PUPPI (GBT/Arecibo)\footnote{\cite{2008SPIE.7019E..1DD, 2017AAS...22913705V}} & 2-8 & 25-800 & 2 & 6.4 \\
	Event Horizon Telescope (2017)\footnote{Per geographic station, J. Weintroub Priv. Comm., cf. \cite{2015PASP..127.1226V}} & 2 & 4000 & 2 & 32  \\
    
	\hline
	\end{tabular}
\end{table*}

\subsection{Closing remarks}

The parameter space within which to search for technological signs of civilizations beyond Earth is vast. The BL \dr ~system described here allows us to search this space orders of magnitude faster and with more sensitivity than has been possible previously. 

We are developing a similar wide-bandwidth recorder system for the Parkes telescope in Australia. Despite the differences between the two telescopes, the Parkes data recorder employs the same architecture and uses much of the same hardware, firmware and software as described here. Output data products will also be comparable, making it straightforward to reuse data analysis tools.

The next generation of radio telescopes will provide exciting opportunities to conduct ever-more powerful searches. For example, the Five-hundred meter Aperture Spherical Telescope (FAST) \citep{2016ASPC..502...93L}, and the MeerKAT 64-element interferometer \citep{2012AfrSk..16..101B}, that are currently nearing their final stages of commissioning, will have unique capabilities that may augment the search. Many aspects of the BL \dr ~system  are directly portable to these  telescopes, offering the prospect for deploying SETI experiments on them quickly and efficiently.

\acknowledgments
Funding for Breakthrough Listen research is sponsored by the Breakthrough Prize Foundation.  We gratefully thank the Breakthrough Listen Advisory Board for their many contributions to the Breakthrough Listen program.  This work made use of the NASA Astrophysics Data System Bibliographic Service.  This work makes use of hardware developed by the Collaboration for Astronomy Signal Processing and Electronics Research (CASPER).

\clearpage

\bibliographystyle{yahapj}
\bibliography{references}

%\appendix
%\section{appendix section}

\end{document}